# Shaped Laser Pulses for Microsecond Time-Resolved Cryo-EM: Outrunning Crystallization During Flash Melting

Constantin R. Krüger[†], Nathan J. Mowry[†], Marcel Drabbels, and Ulrich J. Lorenz[*]

**Affiliation:** Ecole Polytechnique Fédérale de Lausanne (EPFL), Laboratory of Molecular Nanodynamics, CH-1015 Lausanne, Switzerland

† These authors contributed equally

* To whom correspondence should be addressed. Email: ulrich.lorenz@epfl.ch

**Abstract**

Water vitrifies if cooled at rates above $3\cdot10^5$ K/s. Surprisingly, this process cannot simply be reversed by heating the resulting amorphous ice at a similar rate. Instead, we have recently shown that the sample transiently crystallizes even if the heating rate is more than one order of magnitude higher. This may present an issue for microsecond time-resolved cryo-electron microscopy experiments, in which vitreous ice samples are briefly flash melted with a laser pulse, since transient crystallization could potentially alter the dynamics of the embedded proteins. Here, we demonstrate how shaped microsecond laser pulses can be used to increase the heating rate and outrun crystallization during flash melting of amorphous solid water (ASW) samples. We use time-resolved electron diffraction experiments to determine that the critical heating rate is about $10^8$ K/s, more than two orders of magnitude higher than the critical cooling rate. Our experiments add to the toolbox of the emerging field of microsecond time-resolved cryo-electron microscopy by demonstrating a straightforward approach for avoiding crystallization during laser melting and for achieving significantly higher heating rates, which paves the way for nanosecond time-resolved experiments.

If water is cooled at a rate of over 3·10$^5$ K/s,[1] it vitrifies and forms hyperquenched glassy water (HGW), a type of amorphous ice, once the glass transition temperature of 136 K is reached.[2] The successful vitrification of aqueous samples has laid the foundation for cryo-electron microscopy (cryo-EM),[3] which is on its way to become the preferred tool of structural biologists.[4] By outrunning crystallization during the vitrification process, the structure of proteins can be preserved in a frozen-hydrated state. This makes it possible to image them with an electron microscope and use single-particle reconstruction techniques to obtain their three-dimensional structures.[3]

We have recently demonstrated that the vitrification process cannot simply be reversed by heating amorphous ice samples at a similar rate. Instead, partial crystallization occurs even at a heating rate of over 5·10$^6$ K/s.[5] This is a result of the different temperature dependence of the nucleation and growth rates of supercooled water.[2,6] During flash heating, an amorphous ice sample first traverses a temperature range in which fast nucleation occurs before it reaches higher temperatures, at which nucleation largely ceases, but the growth rate surges, so that the sample crystallizes rapidly. In contrast, during hyperquenching, the sequence of events is reversed, which slows crystallization. The critical heating rate at which crystallization can be outrun during flash melting of an amorphous ice sample remains unknown.[5]

The partial crystallization of vitreous ice samples during flash melting[7] also has important implications for microsecond time-resolved cryo-EM experiments, in which a cryo sample is flash melted with a microsecond laser pulse in order to allow protein dynamics to briefly occur while the sample is liquid.[7–9] As the dynamics unfold, the heating laser is switched off, and the sample cools within microseconds and revitrifies, arresting the proteins in their transient configurations, which are subsequently imaged with single particle cryo-EM techniques. Near-atomic resolution reconstructions from revitrified cryo samples demonstrate that the transient crystallization of the sample during laser melting does not alter the structure of embedded particles.[10,11] This is consistent with the observation that high-resolution reconstructions can be obtained from fully devitrified cryo samples.[12] However, it is conceivable that transient crystallization may affect the structure of more fragile proteins and loosely bound complexes or alter their dynamics. It is therefore desirable to avoid crystallization altogether. Here, we show that this can be achieved by using shaped laser pulses with an intense leading edge, which make it possible to dramatically increase the heating rate. Moreover, by systematically varying the heating rate and using time-resolved electron diffraction to probe for crystallization during the melting process, we are able to determine the critical heating rate.

Experiments are performed with a time-resolved transmission electron microscope that we have previously described (Supplementary Methods 1).[13,14] As illustrated in Fig. 1a, we *in situ* deposit a 100 nm thick layer of ASW on a few-layer graphene sheet that is supported by a holey gold film (2 µm holes) on a 600 mesh gold grid (100 K sample temperature). We then flash melt the sample in the center of a grid square with a temporally shaped 30 µs laser pulse (532 nm) and probe whether crystallization occurs by recording a diffraction pattern during the first 10 µs of the melting process with an intense, high-brightness electron pulse[13,14] (Fig. 1b).

Figure 1c schematically illustrates representative shapes of the heating laser pulses, with simulations of the corresponding temperature evolution of the sample shown in Fig. 1d (Supplementary Methods 2). With a simple rectangular pulse (blue), the sample heats up within ~11 µs, before its temperature stabilizes at about 280 K, as previously determined.[5] Once the laser is switched off, the sample cools within a few microseconds and vitrifies.[7] We increase the heating rate during the melting process by adding an initial spike to the laser pulse (red and purple curves). By changing the intensity and duration of this spike while keeping its integral approximately constant, we can adjust the heating rate in the range between $1.6 \cdot 10^7$ and $3.0 \cdot 10^8$ K/s, with the lowest heating rate corresponding to the simple rectangular laser pulse, and the highest rate to a 450 ns spike of more than 17 times the laser power (Supplementary Methods 2). The heating rates we report correspond to averages of the simulated rates between 100 K and 273 K.

Figure 2a shows typical diffraction patterns recorded during the first 10 µs of the flash melting process for heating rates between $1.6 \cdot 10^7$ and $2.2 \cdot 10^8$ K/s. At low heating rates, distinct diffraction spots are visible on top of the broad background of the water diffraction pattern, indicating that the sample has partially crystallized, whereas at higher heating rates, these diffraction features become increasingly fainter. Figure 2b shows the corresponding azimuthally averaged diffraction patterns from the sum of five experiments. These diffraction patterns represent weighted averages of the different temperatures that the sample has explored. At low heating rates, the patterns resemble that of amorphous ice, with a large spacing between the first two diffraction maxima, and exhibit a small contribution from stacking disordered ice. At high heating rates, the sample spends more time at higher temperatures, so that the diffraction pattern increasingly resembles that of stable water. This causes the diffraction intensity to drop and the first diffraction maximum to shift to higher momentum transfer, while the second maximum slightly moves in the opposite direction.[15]

We obtain the critical heating rate by determining the intensity of crystalline features in the diffraction patterns recorded during flash melting (Fig. 2b) as a function of the heating rate. To this end, we decompose the diffraction patterns into the three components shown in Fig. 3a. The diffraction patterns of HGW (green) and liquid water at ~280 K (purple) serve to capture low- and high-temperature structures of liquid water, respectively, while the diffraction pattern of stacking disordered ice (black) is used to describe the crystalline fraction of the sample (Supplementary Methods 3). As shown in Fig. 3b, the experimental diffraction patterns (dashed lines) can be reasonably well described by such weighted sums (solid lines). Figure 3c displays the weight of the three components as a function of the heating rate. As expected, the contribution of the high-temperature structure (purple) increases with heating rate, while the contribution of the low-temperature structure (green) decreases, approaching zero at the highest heating rate. The crystalline component (black, detail in Fig. 3d) has a weight of about 0.11 at the lowest heating rate. This is roughly consistent with our previous estimate that about a third of the sample crystallizes during flash melting with a rectangular laser pulse, if one takes into account that the 10 µs electron pulse probes the sample during the course of the crystallization process. As the heating rate increases, the weight of the crystalline component drops rapidly and approaches zero at a heating rate of about $10^8$ K/s, which we therefore identify as the critical heating rate.

In conclusion, we have determined that during flash melting of ASW samples, the critical heating rate for outrunning crystallization is $10^8$ K/s, more than two orders of magnitude higher than the critical cooling rate during vitrification of about $3 \cdot 10^5$ K/s.[1] Our previous experiments have shown that HGW samples crystallize more rapidly during flash melting than ASW samples, consistent with a 5 times higher nucleation rate.[5] This allows us to estimate that HGW samples should have a 1.5 times higher critical heating rate (Supplementary Methods 4). Note that surface nucleation likely dominates the crystallization process in our thin film samples.[5,16] Therefore, the critical heating rate obtained here represents an upper limit for bulk samples. We similarly expect a lower critical heating rate for typical cryo samples used in microsecond time-resolved cryo-EM experiments, since such samples are usually buffered, and crystallization is known to slow with increasing salt concentration.[1]

Our experiments demonstrate that shaped laser pulses with an intense leading edge provide a straightforward approach for achieving faster heating rates in time-resolved cryo-EM experiments and for avoiding crystallization during flash melting as well as its potentially deleterious effects on particle structure and dynamics. We find that while typical cryo samples transiently crystallize during flash melting with a rectangular laser pulse, crystallization can be comfortably outrun with approximately a 1 µs initial spike of about 9 times the laser power (Supplementary Methods 5). Unless desired, excessive heating of the sample should be avoided by carefully choosing the intensity of the spike based on the heat transfer properties of the sample, which can be characterized experimentally.[15]

Shaped laser pulses with an intense leading edge can also be used to improve the time-resolution of our technique. As we have recently demonstrated, a straightforward strategy for initiating protein dynamics consists in releasing a caged compound[17] while the sample is still in its frozen state.[9] Since the matrix of vitreous ice prevents the embedded proteins from undergoing dynamics, the liberated compound then only becomes active once the sample is liquid. In such experiments, the time resolution is determined both by the heating rate, which determines how fast the sample can be melted and dynamics can be initiated as well as the cooling rate, which dictates how fast the sample can be revitrified and protein motions can be arrested. With a rectangular heating laser pulse, several microseconds can elapse before the sample temperature has stabilized, depending on the heat transfer properties of the sample.[3,5,18] Moreover, if partial crystallization occurs during flash melting, several microseconds more may be required to melt the crystallites that have formed.[5] Particles will therefore be released from their crystalline matrix and start undergoing dynamics at different times, which further lowers the time resolution. As we have shown here, this can be avoided by using shaped laser pulses, which can outrun crystallization and reduce the melting time of the sample to below 600 ns. With suitable hardware, an even shorter melting time can be achieved, so that its contribution to the time resolution becomes negligible. This paves the way for nanosecond time-resolved cryo-EM experiments.

**Acknowledgments:**

The authors would like to thank Dr. Monique S. Straub for her help with the ribosome cryo sample preparation.

**Funding:**

This work was supported by Swiss National Science Foundation Grants PP00P2_163681 and 200020_207842.

**Author contributions:**

Conceptualization: UJL

Methodology: CRK, NJM, UJL

Investigation: CRK, NJM, UJL

Visualization: CRK, NJM, UJL

Funding acquisition: UJL

Project administration: UJL

Supervision: UJL

Writing – original draft: CRK, NJM, UJL

Writing – review & editing: CRK, NJM, MD, UJL

**Competing interests:**

The authors declare that they have no competing interests.

**Data and materials availability:**

All data are available from the corresponding author upon request.

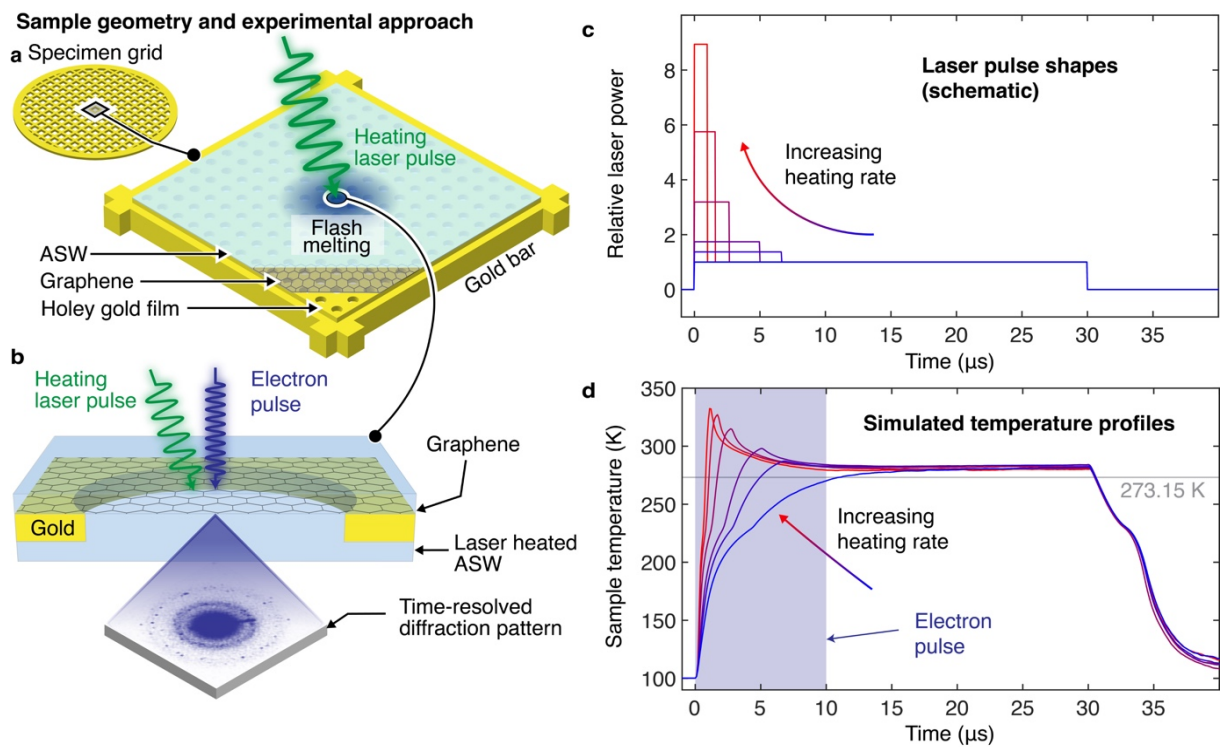

**Figure 1 | Illustration of the experimental approach and simulation of the temperature evolution of the sample. a** Illustration of the sample geometry. A gold mesh supports a holey gold film covered with multilayer graphene, onto which we deposit a 100 nm thick layer of ASW (100 K sample temperature). We then use a shaped microsecond laser pulse to locally melt the sample. **b** We probe for crystallization during the melting process by capturing a diffraction pattern with an intense, 10 μs electron pulse (200 kV accelerating voltage). **c** Schematic illustration of the laser pulse shapes. The heating rate is varied by changing the intensity and duration of the initial spike while keeping its integral constant. **d** Simulation of the temperature evolution of the sample under irradiation with the shaped laser pulses illustrated in **c**. The simulation uses the experimentally determined pulse shapes shown in Supplementary Methods 2. The electron pulse probes the first 10 μs of the melting process.

**Diffraction patterns captured during flash melting**

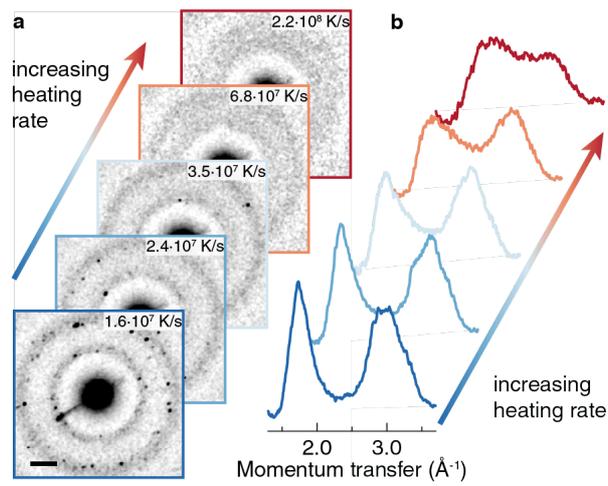

**Figure 2 | Diffraction patterns captured during flash melting. a** Examples of diffraction patterns recorded during the first 10 μs of the melting process for heating rates between $1.6 \cdot 10^7$ and $2.2 \cdot 10^8$ K/s. Scale bar, 1 Å$^{-1}$. **b** Azimuthally averaged diffraction patterns for the heating rates in **a** from the sum of five experiments.

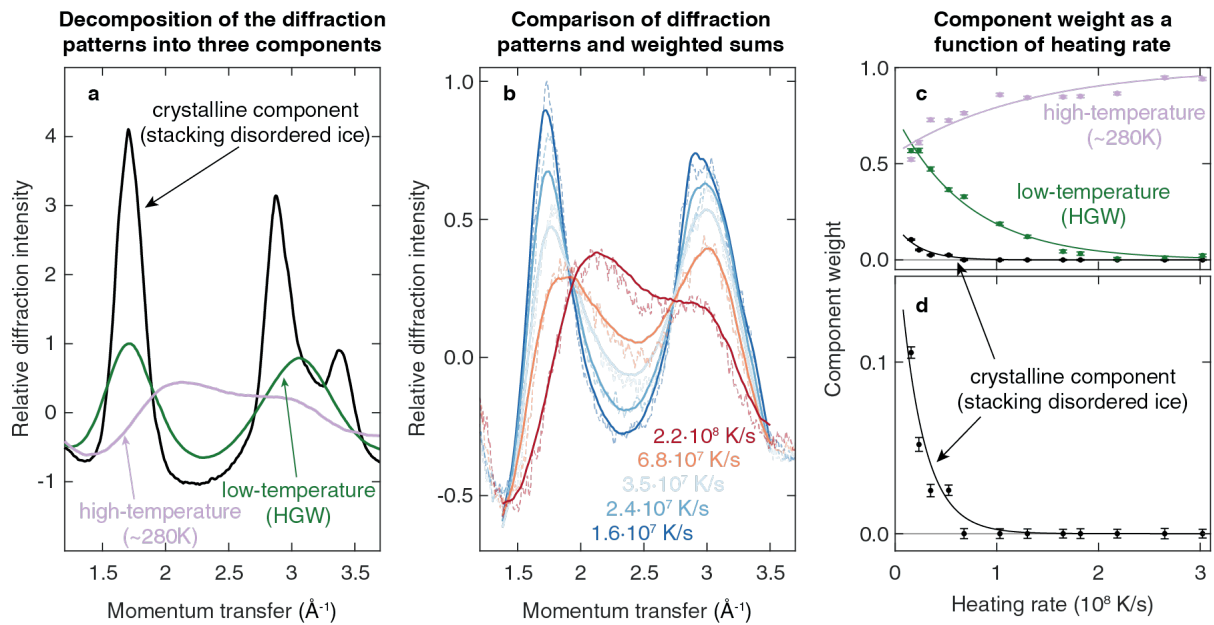

**Figure 3 | Determination of the critical heating rate for outrunning crystallization. a** The diffraction patterns recorded during the melting process (Fig. 2b) can be well reproduced by a weighted sum of the diffraction patterns of a low-temperature component (green, HGW at 100 K), a high-temperature component (purple, water at ~280 K), and a crystalline component (black, stacking disordered ice at 100 K). **b** Weighted sums of the components in **a** (solid lines) show good agreement with the experimental diffraction patterns (dashed lines). **c, d** Weight of the three components in **a** as a function of heating rate. The weight of the crystalline component approaches zero at heating rates exceeding $10^8$ K/s, marking the critical heating rate. Error bars represent standard errors of the fit. The solid lines provide a guide to the eye and are derived from exponential fits.

# Supplementary Information

# Shaped Laser Pulses for Microsecond Time-Resolved Cryo-EM:

# Outrunning Crystallization During Flash Melting

Constantin R. Krüger[†], Nathan J. Mowry[†], Marcel Drabbels, and Ulrich J. Lorenz[*]

**Affiliation:** Ecole Polytechnique Fédérale de Lausanne (EPFL), Laboratory of Molecular Nanodynamics, CH-1015 Lausanne, Switzerland

**This PDF file includes:**

1 Experimental methods

2 Characterization of the laser pulse shapes and determination of the heating rates

3 Decomposition of the diffraction patterns recorded during flash melting into three components

4 Estimate of the critical heating rate for HGW samples

5 Flash melting of a typical cryo sample

[†] These authors contributed equally.

[*] To whom correspondence should be addressed. E-mail: ulrich.lorenz@epfl.ch

# 1 Experimental methods

Experiments are performed with a JEOL 2010F transmission electron microscope that we have modified for time-resolved experiments.(*1*, *2*) Sample supports are prepared by transferring 6-8 layer graphene onto Quantifoil (Au) R2/1 (N1-A15nAu60-50) specimen grids (50 nm thick holey gold film with 2 µm diameter holes, 1 µm apart on 600 mesh gold).(*3*, *4*) The sample is held at a temperature of 100 K, and a 100 nm thick layer of ASW is deposited *in situ* as previously described.(*3*, *4*) Here, the cold shield surrounding the sample is cooled to liquid nitrogen temperature in order to allow for faster deposition rates. The time-resolved electron diffraction experiments follow the same sequence of events as before.(*3*, *4*) Static diffraction patterns of the sample are captured before and after each experiment, and a time-resolved diffraction patterns is captured during the first 10 µs of the laser melting process to probe for crystallization. After the experiment, the sample is evaporated, and a fresh ASW sample is deposited. The diffraction patterns are analyzed as previously described and are shown with the diffraction background subtracted.(*3*, *4*)

# 2 Characterization of the laser pulse shapes and determination of the heating rates

In order to systematically vary the heating rate, we change the amplitude and duration of the initial spike of the shaped laser pulse, while keeping the integral of the spike approximately constant. Figure S1 shows the experimental laser pulse shapes, as recorded with a photodiode (averages of 10). The area under the initial spike varies by less than 10 % across the different pulse shapes.

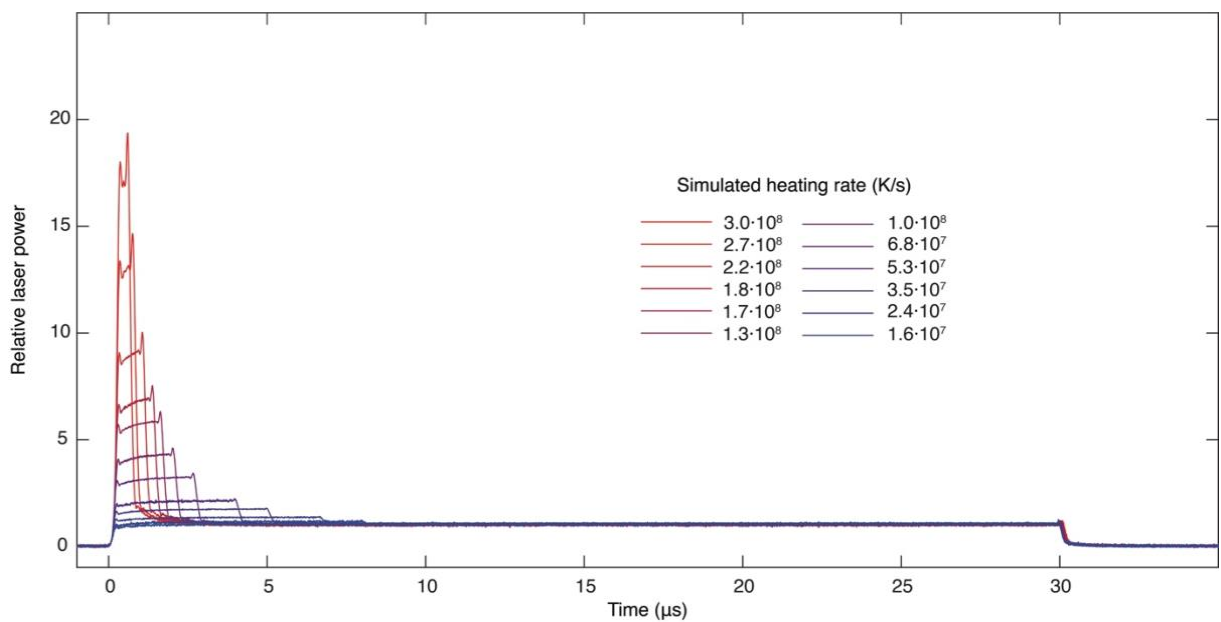

**Figure S1. Experimental laser pulse shapes as determined with a photodiode.** The simulated heating rates are listed for each pulse shape.

The amplitude and duration of the spike are listed in Table 1 for each laser pulse shape, together with the heating rate, as determined from heat transfer simulations. The temperature evolution of the sample is simulated with COMSOL Multiphysics 6.1, with the simulation parameters as previously described.(*3*, *4*) We simulate the experimentally determined laser pulse shape and adjust the simulated laser power such that at the end of a rectangular laser pulse, the sample temperature plateaus at 280 K, as previously determined in a similar experiment.(*4*) Within the estimated error of this plateau temperature of less than ~10 K, the heating rate changes only marginally. We report the average heating rate of the sample between 100 K and 273 K in the volume probed by the electron pulse (~1.5 µm beam diameter). The largest temperature difference across this volume occurs at the highest heating rate and briefly reaches about 50 K.

| Relative spike amplitude | Spike duration (µs) | Simulated heating rate (K/s) |
|---|---|---|
| 17.1 | 0.45 | $3.0 \cdot 10^8$ |
| 13.0 | 0.64 | $2.7 \cdot 10^8$ |
| 8.94 | 0.97 | $2.2 \cdot 10^8$ |
| 6.74 | 1.30 | $1.8 \cdot 10^8$ |
| 5.75 | 1.59 | $1.7 \cdot 10^8$ |
| 4.20 | 2.00 | $1.3 \cdot 10^8$ |
| 3.19 | 2.69 | $1.0 \cdot 10^8$ |
| 2.14 | 4.02 | $6.8 \cdot 10^7$ |
| 1.74 | 5.09 | $5.3 \cdot 10^7$ |
| 1.37 | 6.68 | $3.5 \cdot 10^7$ |
| 1.16 | 7.99 | $2.4 \cdot 10^7$ |
| 1.00 | — | $1.6 \cdot 10^7$ |

**Table 1. Parameters of the shaped laser pulses and simulated heating rates.** The amplitude of the initial spike is indicated in multiples of the laser power of a simple rectangular pulse. Both the amplitude and duration are measured with a photodiode.

**3 Decomposition of the diffraction patterns recorded during flash melting into three components**

In order to determine whether crystallization occurred during flash melting, we decompose the diffraction patterns captured during the first 10 µs of the melting process into a high- and a low-temperature component as well as a crystalline component. The weights of the components are determined with least-squares minimization.

For the low-temperature component, we use the diffraction pattern of the HGW sample that is obtained after each laser pulse, where we have averaged diffraction patterns from 73 individual experiments, each recorded with 20 boosted electron pulses of 10 µs duration, fired at 10 Hz repetition rate.

For the high-temperature component, we use the diffraction pattern of liquid water (~280 K), which we record at the end of the laser pulse with a 10 µs boosted electron pulse. The diffraction pattern in Fig. 3a is an average of 4 patterns, smoothed with a rolling average filter.

For the crystalline component, we use the diffraction pattern of a devitrified sample, which predominantly consists of stacking disordered ice (20 boosted electron pulses of 10 µs duration, fired at 10 Hz repetition rate). The sample was prepared by irradiating a freshly deposited ASW sample with 10 laser pulses of 30 µs duration, with the laser power reduced to two thirds of the power needed for achieving melting and revitrification. At this laser power, the sample temperature plateaus in the deeply supercooled regime, and devitrification occurs.

**4 Estimate of the critical heating rate for HGW samples**

We have previously shown that HGW samples exhibit faster crystallization kinetics during flash melting than ASW samples, a difference that can be explained if one assumes that the nucleation rate of HGW is about 5 times as large as that of ASW.(*4*) This allows us to estimate the critical heating rate for HGW samples with the help of simulations of the crystallization kinetics. Our model, which we have detailed previously,(*4*) is based on classical nucleation and growth theory(*5*, *6*) and uses experimental nucleation and growth rates.(*7*, *8*) In the deeply supercooled regime, where the nucleation rate goes through a maximum, experimental values are largely not available.(*7*) Moreover, in our thin film samples, surface nucleation likely plays an important role, which increases the nucleation rate.(*4*) We therefore treat the maximum nucleation rate of deeply supercooled water as a free parameter.

We begin by simulating the flash melting of an ASW sample with a constant heating rate of $1 \cdot 10^8$ K/s, which corresponds to the critical heating rate that we have determined experimentally. We then adjust the maximum nucleation rate of deeply supercooled water such that only a small fraction of the sample of 0.1 % has crystallized when the sample reaches the melting point. In order to simulate the flash melting of HGW, we then increase the maximum nucleation rate by a factor of 5, which increases the crystalline fraction of the sample when it reaches the melting point to 0.5 %. We find that in order to maintain the crystalline fraction at 0.1 %, the heating rate has to be increased by a factor of 1.5. This allows us to estimate that the critical heating rate for HGW samples is 1.5 times higher than for ASW.

Note that this estimate is not sensitive to our choice of the crystalline fraction that is observed at the critical heating and barely changes if we reduce the crystalline fraction by one or two orders of magnitude.

## 5 Flash melting of a typical cryo sample

While typical cryo samples transiently crystallize during flash melting with a rectangular laser pulse, crystallization can be outrun with a shaped laser pulse is used. This is shown in Fig. S2 for a cryo sample of the 50S subunit of the ribosome (20 mM HEPES at pH 7.5, 100mM NaCl, 2mM $MgCl_2$) on an UltrAuFoil R1.2/1.3 (N1 - A14nAu30 – 01) specimen grid (50 nm thick holey gold film with 1.2 µm diameter holes, 1.3 µm apart on 300 mesh gold). As shown in Fig. S2a, when the sample is flash melted with a rectangular 30 µs laser pulse, transient crystallization is evident in a diffraction pattern recorded during the melting process (captured with a 10 µs electron pulse between 3 µs and 13 µs). In contrast, crystallization can be outrun with an initial spike of 0.94 µs duration and 8.94 times the laser power (Fig. S2b, diffraction pattern recorded between 0 µs and 10 µs).

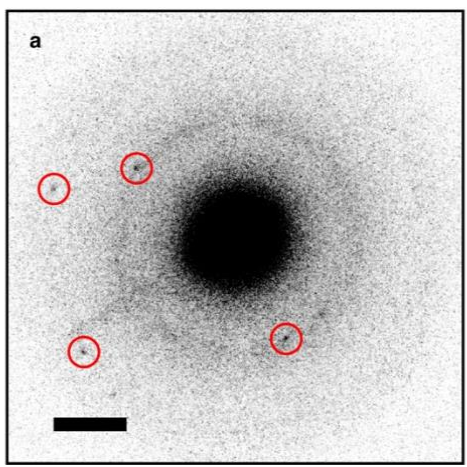
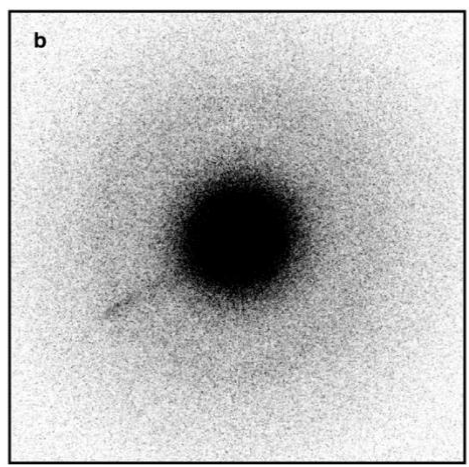

**Figure S2. Diffraction patterns of a typical cryo sample recorded during flash melting with a rectangular and a shaped laser pulse. a** Transient crystallization is evident during flash melting with a rectangular 30 µs laser pulse (crystalline diffraction features highlighted with circles). The time-resolved diffraction pattern was captured with a 10 µs electron pulse between 3 µs and 13 µs.

**b** Crystallization can be outrun with a shaped laser pulse (0.94 µs initial spike with 8.94 times the laser power). The time-resolved diffraction pattern was recorded between 0 µs and 10 µs. Scale bar, 1 Å$^{-1}$.